\documentclass[journal=nalefd,manuscript=letter,email=false]{achemso}

\usepackage{chemformula} 
\usepackage[T1]{fontenc} 

\usepackage{siunitx}               
\usepackage{changes}
	
	\title{Optical transparency induced by a largely Purcell-enhanced quantum dot in a polarization-degenerate cavity}

\author{Harjot Singh}
\affiliation{Department of Electrical and Computer Engineering, Institute for Research in Electronics and Applied Physics, and Joint Quantum Institute, University of Maryland, College Park, MD 20742, USA}
\alsoaffiliation{These authors contributed equally to this work}

\author{Demitry Farfurnik*}

\affiliation{Department of Electrical and Computer Engineering, Institute for Research in Electronics and Applied Physics, and Joint Quantum Institute, University of Maryland, College Park, MD 20742, USA}
\alsoaffiliation{These authors contributed equally to this work}
\alsoaffiliation{Corresponding Author, dimka$\_$f13$@$yahoo.com}

\author{Zhouchen Luo}
\affiliation{Department of Electrical and Computer Engineering, Institute for Research in Electronics and Applied Physics, and Joint Quantum Institute, University of Maryland, College Park, MD 20742, USA}

\author{Allan S. Bracker}
\affiliation{Naval Research Laboratory, 4555 Overlook Avenue SW, Washington, D.C., 20375, USA.}

\author{Samuel G. Carter}
\affiliation{Naval Research Laboratory, 4555 Overlook Avenue SW, Washington, D.C., 20375, USA.}
\alsoaffiliation{Present address: Laboratory for Physical Sciences, University of Maryland, College Park, MD 20740, USA}

\author{Edo Waks}
\affiliation{Department of Electrical and Computer Engineering, Institute for Research in Electronics and Applied Physics, and Joint Quantum Institute, University of Maryland, College Park, MD 20742, USA}

\begin{document}

\begin{abstract}
Optically-active spin systems coupled to photonic cavities with high cooperativity can generate strong light-matter interactions, a key ingredient in  quantum networks. But obtaining high cooperativities for quantum information processing often involves the use of photonic crystal cavities that feature a poor optical access from the free space, especially to circularly polarized light required for the coherent control of the spin.  Here, we demonstrate coupling with cooperativity as high as $8$ of an InAs/GaAs quantum dot to a fabricated bullseye cavity that provides nearly degenerate and Gaussian polarization modes for efficient optical accessing. We observe spontaneous emission lifetimes of the quantum dot as short as $80$ ps (a $\approx 15$ Purcell enhancement) and a $\approx 80\%$ transparency of light reflected from the cavity. Leveraging the induced transparency for photon switching while coherently controlling the quantum dot spin could contribute to ongoing efforts of establishing quantum networks.
 
\end{abstract}
	\maketitle
\section*{}

In recent years, optically-active quantum dots have emerged as useful resources for photonic quantum technologies. Quantum dots emit single photons with high brightness and indistinguishability \cite{Santori2001,Gazzano2013,Ding2016,Somaschi2016,Liu2018,Scholl2019,Wang2019,Tomm2021}, which makes them promising as sources of single and entangled photons for photonic quantum computing \cite{Schwartz2016,Huber2017,Segovia2019,Istrati2020}. In addition, these dots can be electrically charged with a single electron or a single hole, thereby offering a ground state spin qubit \cite{Bayer2002,Press2008,Carter2013,Warburton2013,Bechtold2015,Stockill2016,Farfurnik2021}. Strongly coupling a quantum dot spin to a photonic cavity could provide an interface between a single photon and a single spin for quantum information processing \cite{Sun2018,Najer2019}, thereby contributing to the ongoing efforts of establishing quantum networks \cite{Kimble2008,Lodahl2018}. Such strong coupling requires a sufficiently high cooperativity between the spin and the cavity, which typically involves the use of high-Q ($>10,000$) cavities \cite{Sun2018,Najer2019}. An ultrahigh cooperativity has been recently achieved between a quantum dot and a high-Q tunable microcavity formed by utilizing the advanced fabrication of convex mirrors \cite{Najer2019}. An alternative approach for achieving such high cooperativities utilizes simple nanofabrication tools (e.g., electron beam lithography) to fabricate photonic crystal cavities. However, high-Q photonic crystal cavities often feature poor optical access to external light from the free space due to their divergent far field emission patterns \cite{Carter2013,Sun2018,Luo2019}. This poor access limits the ability to optically excite and collect photons emitted from quantum dots, as well as to coherently control the quantum dot spin, which requires circularly polarized light \cite{Press2008,Carter2013,Stockill2016,Farfurnik2021}.

To efficiently interface quantum dots with light often involves their coupling to low-Q ($<1000$) cavities such as gratings and micropillars  \cite{Davanco2011,Lodahl2015,Ding2016,Somaschi2016,Liu2018,Wang2019,Liu2019,Kolatschek2019,Moczala2020,Kolatschek2021}.  Low-Q cavities can increase the optical density of states in the environment of a quantum dot, thereby Purcell-enhancing \cite{Purcell1946} the rate of spontaneous emission of single photons from the dot. In addition, cavities that provide Gaussian far field emission patterns can improve the efficiency of exciting and collecting photons from the quantum dot via confocal optical setups. For example, circular gratings formed by the  periodical etching of rings from a substrate material ("bullseye" cavities) have been used to optically interface single defects in diamond \cite{Li2015} and quantum dots \cite{Davanco2011,Sapienza2015,Wang2019,Liu2019,Kolatschek2019,Moczala2020,Kolatschek2021}.  In particular, InAs/GaAs quantum dots produced optical emission of single photons with lifetimes of $\approx 200$ ns \cite{Moczala2020}. Improving the efficiency of collecting photons from quantum dots coupled to such low-Q cavities can be achieved by introducing ellipticity to the structure\cite{Wang2019,Tomm2021}, but this prevents the access to the cavity using circularly polarized light.  Another downside of low-Q cavities is their high loss of photons, which may result in low spin-cavity cooperativities, thereby significantly limiting the performance of the cavities for quantum networking \cite{Sun2018,Najer2019}. To date, a low-Q cavity that provides a high cooperativity spin-photon interface with an efficient, polarization independent, optical access has yet to be demonstrated.

Here, we efficiently couple InAs/GaAs quantum dots embedded in a charge tunable device (a p-i-n-i-n diode) \cite{Vora2015,Lobl2017,Luo2019} to low-Q ($\approx 1000$) bullseye cavities with nearly degenerate polarization modes.  By leveraging the low charge noise associated with the device, we measure spontaneous emission lifetimes of quantum dots as short as $\approx$ 80 ps  (a Purcell enhancement of $\approx$ 15), which are more than two times shorter than  previously observed for InAs/GaAs quantum dots in nearly degenerate bullseye cavities \cite{Moczala2020}, and are close to the state-of-the-art lifetimes of such dots in microcavities \cite{Somaschi2016,Ding2016,Wang2019,Tomm2021}. By measuring a dip in the reflected light from a bullseye cavity caused by its coupling to an uncharged quantum dot, we extract a cooperativity of $\approx 8$ between the cavity and the dot, which highlights the potential of the bullseye cavities as spin-photon interfaces. Combined with the enhanced efficiencies of optically exciting the quantum dot spin and collecting the emitted photons, the fabricated bullseye cavities offer a promising platform for quantum information processing utilizing electrically charged quantum dots.

We perform measurements on InAs quantum dots embedded in GaAs, deterministically charged by applying a DC bias voltage on a p-i-n-i-n diode [Fig. 1(a)] \cite{Lobl2017,Luo2019} (see Section I of Supporting Information). Under the application of an external magnetic field of $B=9$ T perpendicular to the sample growth axis (Voigt geometry), we observe Zeeman splitting of a single optical transition of the dots into two and four transitions for uncharged and electrically charged quantum dots, respectively \cite{Bayer2002} (see Section II of Supporting Information). For electrically charged quantum dots, working in the Voigt geometry provides efficient paths to optically initialize the quantum dot spin on nanosecond timescale and to coherently control it utilizing a Lambda system \cite{Press2008,Stockill2016,Farfurnik2021}. Our measurements of six different optical transitions of quantum dots in the bulk indicate their photon emission rates on timescales of 1.1-1.3 ns, consistent with previous observations \cite{Sapienza2015}. By analyzing these statistical results, we extract the typical lifetime of
dots in our sample of $1280 \pm 60$ ps.

To improve the optical interface of the quantum dots, we fabricate bullseye cavities by etching rings from the semiconductor membrane consisting of the p-i-n-i-n diode and a sacrificial AlGaAs-Si layer [white areas in Fig. 1 (a)] (see Section I of Supporting Information). Figure 1 (b) shows the SEM image of a typical fabricated structure, where the dark areas represent the etched material and the vertical and horizontal tapered lines represent "bridges" that enable the electrical charging of the dots and prevent the structure from collapsing \cite{Li2015}. A finite-difference-time-domain simulation of the far field emission of light from the cavity results in a nearly Gaussian pattern with waist smaller than the numerical aperture (0.68) of collecting photons via the objective lens of our optical setup, thereby highlighting the potential of the cavity in efficiently exciting and collecting photons from quantum dots coupled to it. In addition, the simulated far field emission modes are polarization degenerate, which we can experimentally verify by rotating quarter and half waveplates in the collection path of the setup (see Fig. S2 of Supporting Information). Such polarization degeneracy positions bullseye cavities promising for the coherent control of the spin of electrically charged quantum dots coupled to the cavity, which typically requires circularly polarized light \cite{Press2008,Stockill2016,Farfurnik2021}.

 \begin{figure}[h]
 	\centering
 	\includegraphics[width=0.45\textwidth]{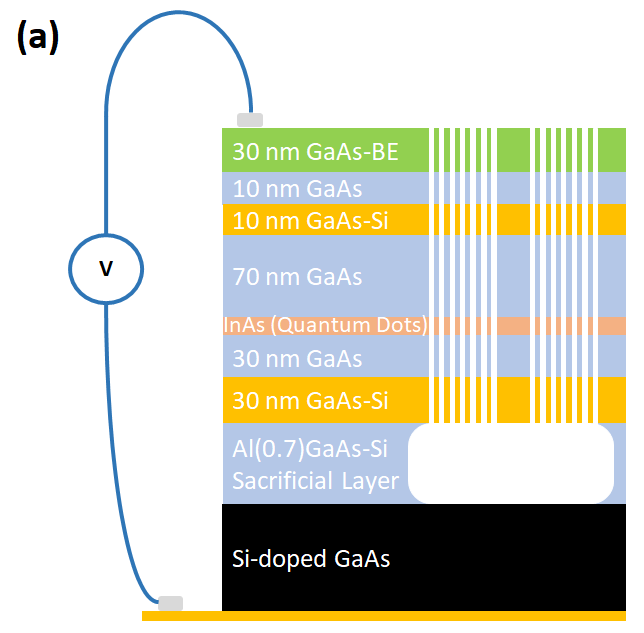}\\
 	\includegraphics[width=0.27\textwidth]{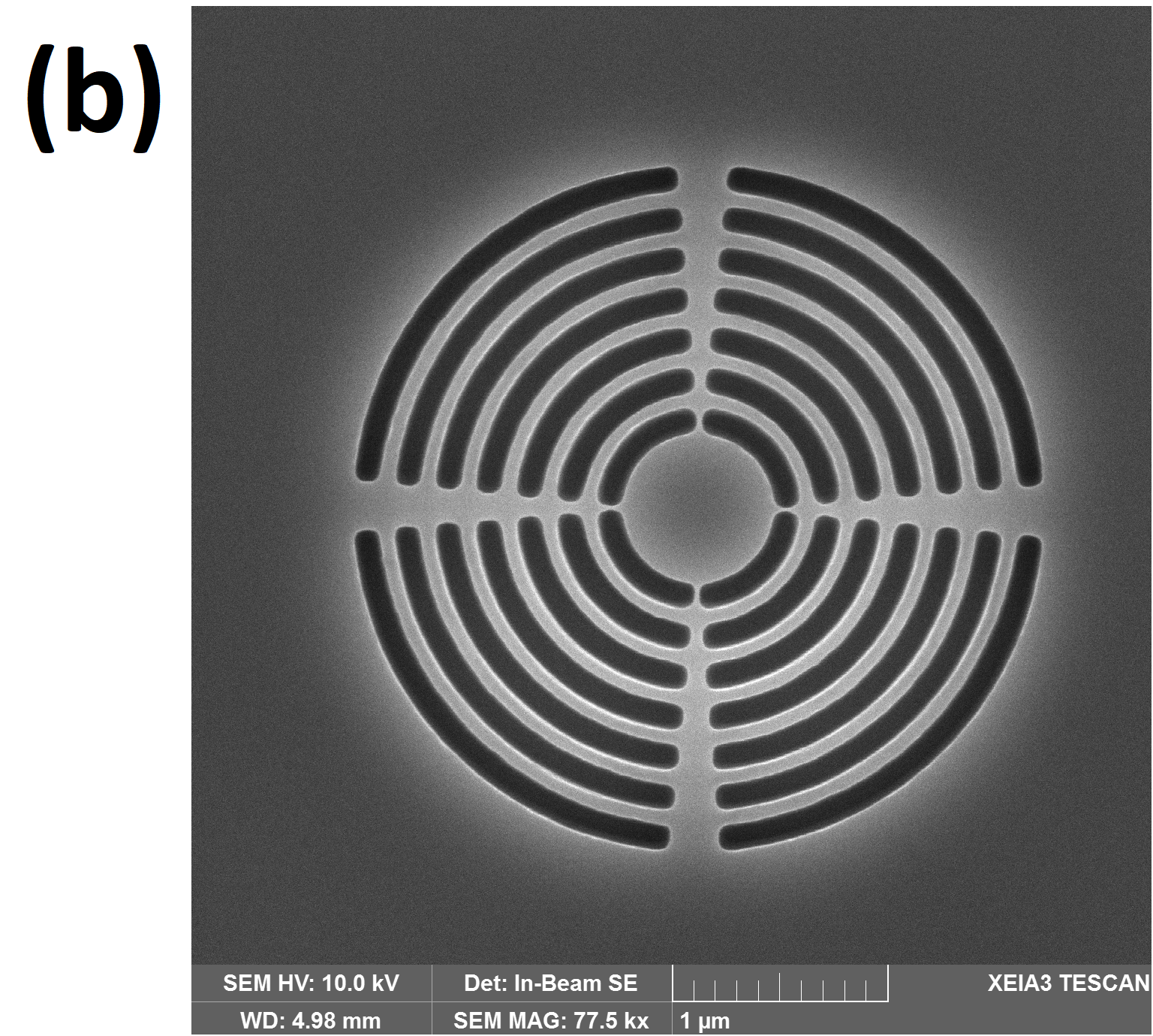}
 	\includegraphics[width=0.33\textwidth]{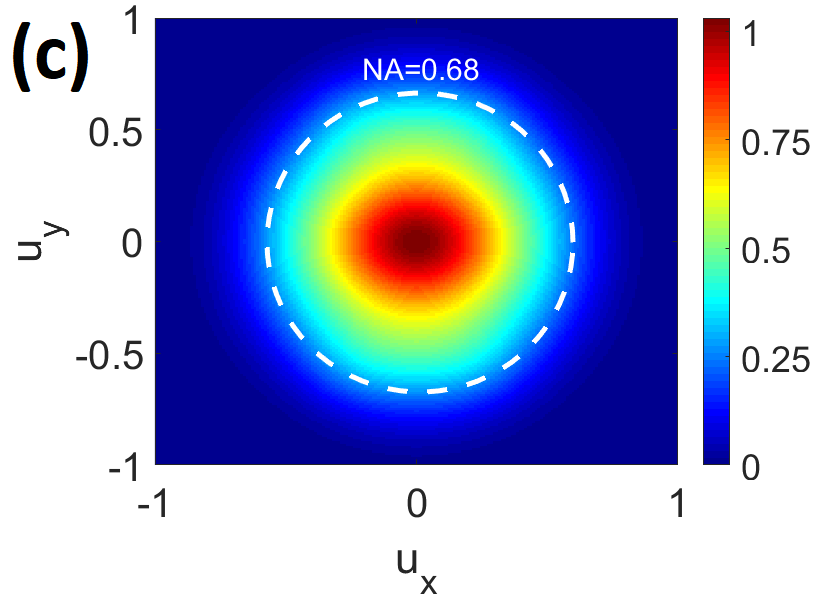}
 	\caption{(a) A cross section of a semiconductor device consisting of a p-i-n-i-n diode for the deterministic charging of quantum dots within a thin InAs layer. The white areas illustrate the etched material for the fabrication of a bullseye cavity. (b) An SEM image of a bullseye cavity fabricated on a GaAs sample. The seven dark rings represent areas etched from the sample.   (c) A finite-difference-time-domain simulation of the far field emission pattern of the bullseye cavity. The polarization independent Gaussian emission pattern matches the numerical aperture of our objective lens (dashed white line), thereby indicating the efficient interfacing of light with single photon emitters coupled to the cavity.}
 	\label{fig:fig1}
 \end{figure}
 
To study the impact of the bullseye cavities on the optical properties of quantum dots, we locate a cavity coupled to two separate uncharged dots [Fig. 2 (a)]. By applying above-band laser light (an $\approx$ 860 nm non-resonant excitation), we generate charge carriers in the wetting layer of the sample that induce spontaneous photon emission from the cavity [solid blue line in Fig. 2 (a)]. The measured quality factor of $\approx 1070$ of the cavity agrees with our theoretical predictions from finite-difference-time-domain simulations. After the observation of photon emission from the cavity, we tune the voltage of the diode to the bias plateau where two quantum dots are neutral. We identify the spontaneous emission of photons from the dots [dashed red lines in Fig. 2 (a)] by reducing the power of the above-band laser to eliminate the photoluminescence from the cavity. Sweeping the external magnetic field from $B=0$ to $B=9$ T (see Section II of Supporting Information) verifies that these dots are uncharged, with one dot (labeled "BE, Dot 1")  on spectral resonance with the central frequency of the cavity, and the other (labeled "BE, Dot 2") slightly detuned from this frequency.

\begin{figure}[h]
	\centering
	\includegraphics[width=0.45\textwidth]{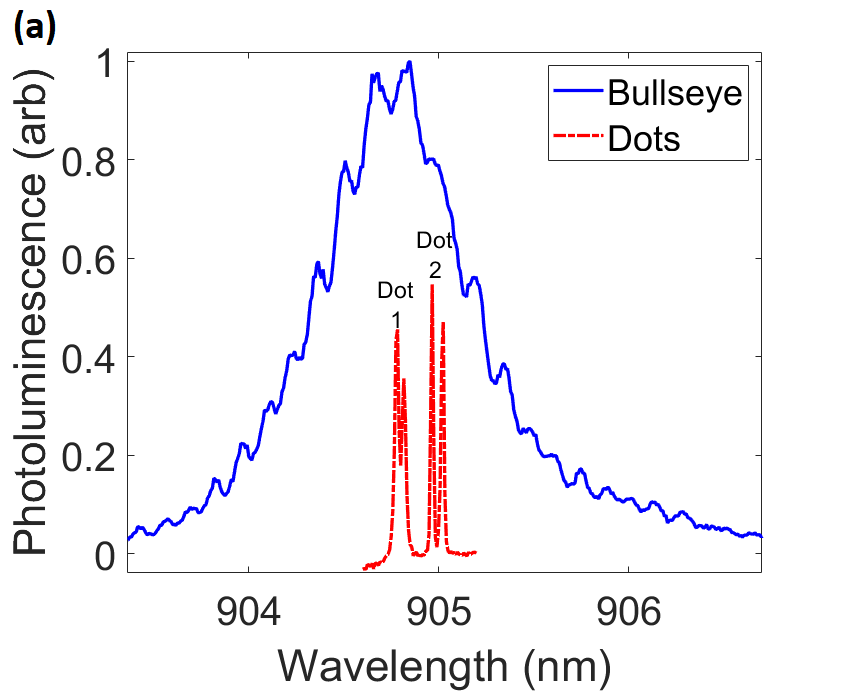}
		\includegraphics[width=0.45\textwidth]{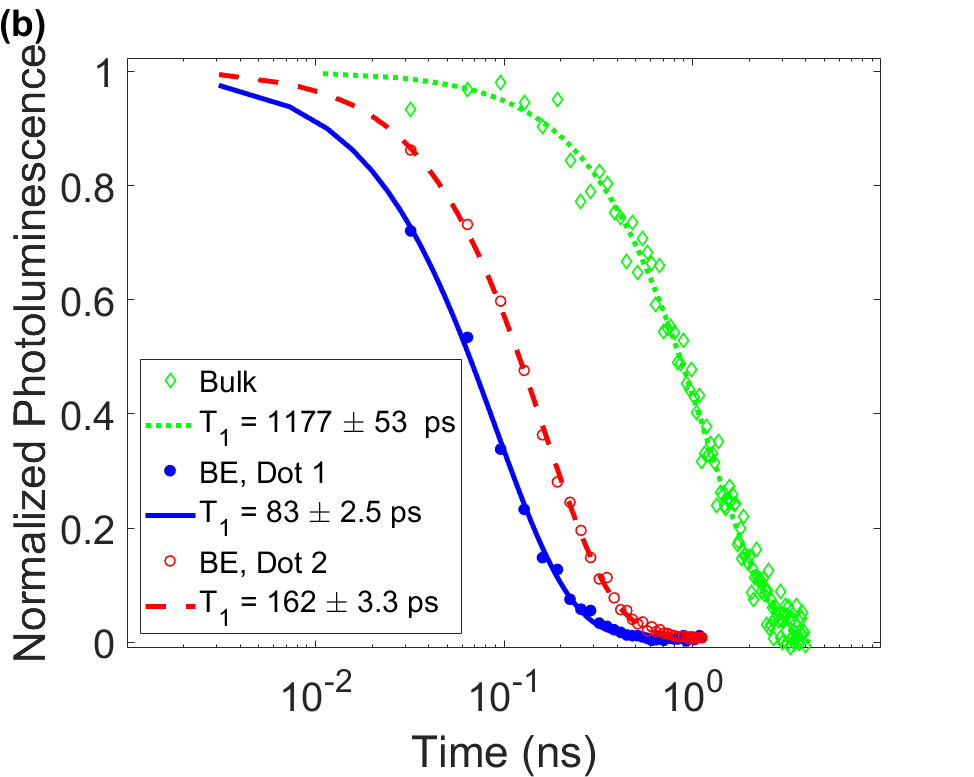}
	\caption{(a) Photoluminescence spectra of a bullseye cavity (solid blue line) and two uncharged quantum dots coupled to the cavity (dashed red line). (b) Time-resolved measurements of the optical lifetimes of quantum dots in bulk (dotted green line), as well as of two quantum dots coupled to a bullseye cavity labeled in (a) as "Dot 1" (solid blue line) and "Dot 2" (dashed red line).}
	\label{fig:fig2}
\end{figure}

Figure 2 (b) compares the spontaneous emission lifetimes of the two dots in the bullseye cavity with the lifetime of a quantum dot in the bulk. In the presence of the cavity, the emission lifetime shortens from $\approx$ 1.2 ns [e.g., the dotted green line in Fig. 2 (b)] for bulk dots down to $\approx$ 160 ps (a Purcell enhancement of $7.93\pm 0.41$ compared to the statistically calculated lifetime in the bulk) for the dot spectrally detuned from the cavity, and down to  $\approx$ 80 ps (a calculated Purcell enhancement of $15.35\pm 0.85$) for the dot resonant with the cavity. The latter is close to the state-of-the-art lifetimes measured on InAs/GaAs quantum dots embedded in microcavities \cite{Ding2016,Somaschi2016,Wang2019,Tomm2021}, and is shorter by a factor of $\approx 2$ than those previously measured on InAs/GaAs quantum dots in bullseye cavities with nearly degenerate polarization modes \cite{Moczala2020}. We attribute this improvement to two main factors. First, our fabrication of bullseye structures involves the etching of a sacrificial layer below the cavity, whereas the rings of the bullseye structures in previous fabrications  \cite{Davanco2011,Sapienza2015} either were not etched all the way down to this layer or incorporated a metallic mirror at the bottom\cite{Moczala2020}. While partial etching or the addition of a metallic mirror could improve the collection efficiency of photons scattered from the sample, these procedures may reduce the quality factor of the bullseye cavity, thereby providing a smaller Purcell enhancement compared to the ones we observe for suspended structures. The second factor that contributes to the enhanced optical emission rate is the reduced charge noise associated with the p-i-n-i-n diode \cite{Lobl2017,Luo2019}, which minimizes the effects of spectral wandering of quantum dots coupled to the cavity that may degrade the Purcell enhancement of their emission.  

To study the potential of bullseye cavities as spin-photon interfaces, we measure the reflectivity of light from a cavity coupled to an uncharged quantum dot in the absence of external magnetic field. We sweep the frequency of a weak ($\approx 0.5$ nW) continuous-wave laser and measure the intensity of the signal reflected from the cavity [blue dots in Fig. 3 (a)]. A dip in the cavity reflectivity with a contrast of $\approx 80\%$ emerges at the wavelength of the optical transition of the quantum dot. This optical transparency is caused by the destructive interference between two optical transitions with opposite phases generated due to the coupling between the cavity and the quantum dot (in analogy with electromagnetically induced transparency where two transitions in an atomic three level system interfere destructively) and improves with the Purcell factor \cite{Waks2006}. The observed reflectivity pattern is asymmetric, namely a Fano resonance \cite{Fano1961,Rybin2009,Limonov2017}, possibly due to a small splitting  between the cavity polarization modes or to an additional interference effect resulting from the membrane of our sample (see Section III of Supporting Information). The measured reflectivity agrees with simulation results based on a theoretical Jaynes-Cummings model \cite{Englund2005,Luo2019} [red line in Fig. 3 (a)] considering coupling strength of $g= 35$ GHz between the quantum dot and the cavity and photon losses from the cavity and the dot of $\kappa = 310$ GHz and $\gamma= 1$ GHz, respectively (see Section III of Supporting Information). The relatively high cooperativity between the dot and the cavity extracted from the model, $C=\frac{2g^2}{\kappa\gamma} \approx 8$, suggests that bullseye cavities can be used for photon switching and for interfacing single photons with single spins \cite{Sun2018,Najer2019}.

\begin{figure}[h]
	\centering
	\includegraphics[width=0.45\textwidth]{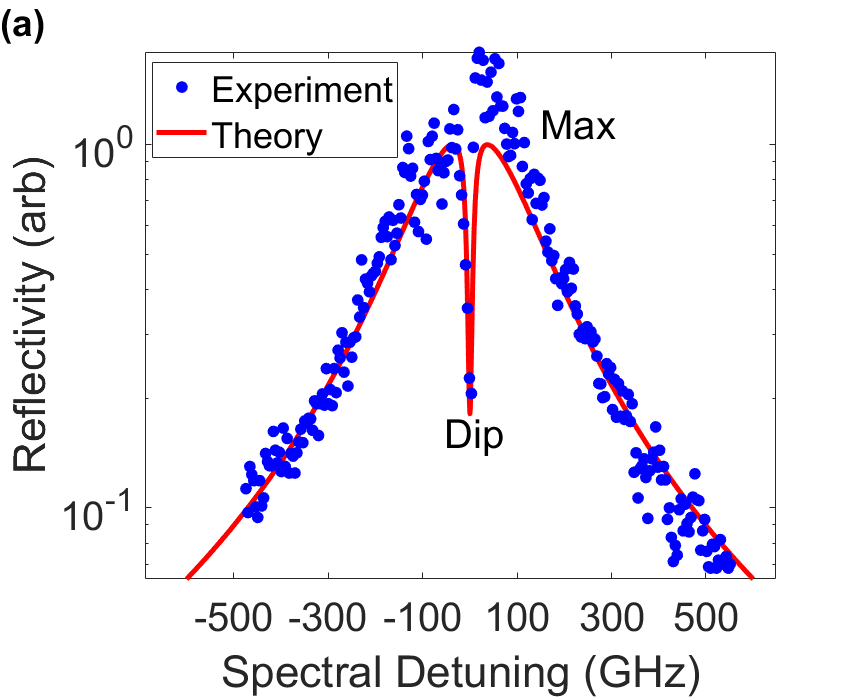}
	\includegraphics[width=0.45\textwidth]{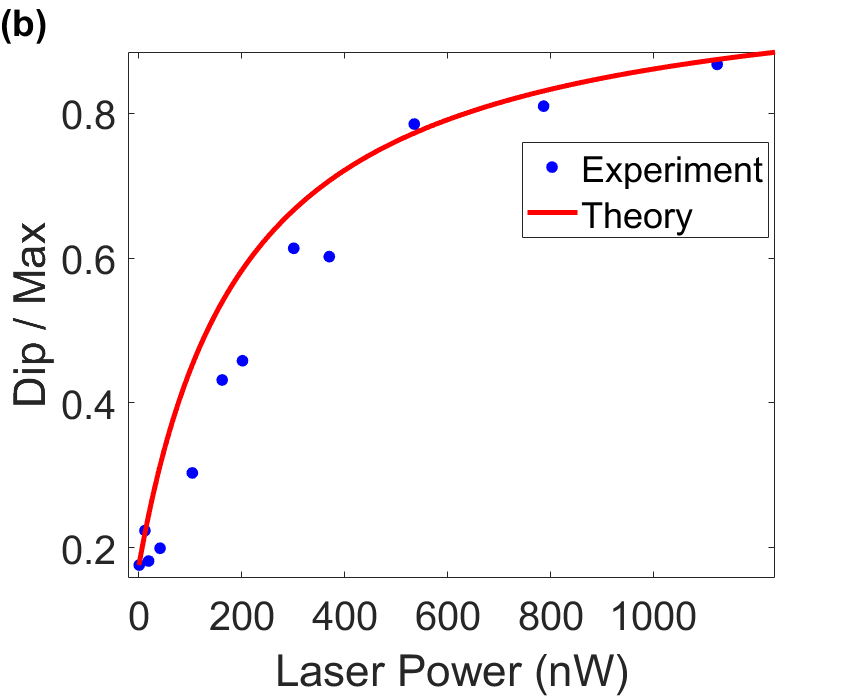}
	\caption{(a) The intensity of laser light reflected from a bullseye cavity coupled to an uncharged quantum dot ($B = 0$) as a function of the spectral detuning of the laser from the optical transition of the dot. The blue dots represent experimental results and the dashed red line represents simulation results considering a Jaynes-Cummings model. The quantum dot optical transition induces transparency of the cavity reflectivity, which can be used for photon switching. (b) The intensity of reflected at the cavity dip coupled to a quantum dot (marked "Dip" in (a)) normalized by the reflectivity of a bare cavity (marked "Max" in (a)) as a function of the incident laser power. The blue dots represent experimental results and the dashed red line represents simulation results considering an efficiency of $8\%$ of optically accessing the cavity from free space optics.}
	\label{fig:fig3}
\end{figure}

We can estimate the efficiency of coupling photons to quantum dots in such interfaces by plotting the intensity of light reflected at the cavity dip as a function of the incident laser power [Fig. 3 (b)]. Fitting the experimentally measured dip reflectivity trend to the theoretically simulated one \cite{Englund2005} reveals that $\approx 8\%$ of the incident light reaches the cavity (see Section III of Supporting Information). This efficiency is much greater than the ones we observe in the same experimental setup utilizing photonic crystal (e.g., L3) cavities, indicating that the far field emission pattern of the bullseye cavity is concentrated at small angles, as expected from simulations [Fig. 1(c)]. The main factor limiting the coupling efficiency is the mismatch between the numerical aperture of the objective lens in our setup (0.68) and the angle of the cavity far field emission mode corresponding to a $1/e$ relative intensity $(\approx 0.36)$ [Fig. 1 (c)]. This mismatch leads to a factor $\approx 3.6$ degradation in the efficiency that can be avoided by changing the lenses in our experimental setup. After additionally growing a distributed Bragg reflector at the bottom of the sample the sample to  act as a mirror, we expect to optically access quantum dots in next generation bullseye cavities with efficiencies of over $60\%$. Beyond improving the rates of optical excitation and photon collection, such efficient access of light may enable multi-pulse coherent control of quantum dot spins in cavities \cite{Farfurnik2021}.     

We examine a bullseye cavity coupled to such a spin, namely a single electron spin qubit confined in a ("charged") quantum dot. Under an external magnetic field of $B = 9$ T, the application of a series of ultrashort above-band laser pulses (i.e., they are much shorter than the optical emission rates of the dot) reveals the optical transitions of the dot [Fig. 4 (a)]. Compared to dots in the bulk, the laser power required for the saturation of the photoluminescence signal from the quantum dot in the cavity is an order of magnitude weaker, and the intensity of this signal is $\approx$ 25 stronger. Optically accessing the device should be even more efficient for light in spectral resonance with the cavity, namely spectrally detuned by $\sim$ 1 nm from the optical transitions of the dot.
Here, due to the observed spectral detuning, the Purcell enhancement of photon emission via the optical transitions of the examined electrically charged dot of $[\approx 8-9$, inset of Fig. 4 (a)] is smaller than the one observed for the uncharged dot [Fig. 2(b)]. However, this spectral detuning can be leveraged to boost the efficiency of optical pulses that coherently control the quantum dot spin applied in resonance with the cavity, as such pulses must be spectrally detuned from the dot's optical transitions \cite{Press2008,Stockill2016,Farfurnik2021}. Another observation that highlights the potential of bullseye cavities for boosting spin coherent control is the emission of photons from all four optical transitions of the dot. As illustrated by Fig. 4 (a), the charged quantum dot emits photons at polarizations orthogonal to each other, with directions dictated by the external magnetic field. The observation of photoluminescence from all four transitions is consistent with the expected polarization degeneracy of the cavity mode (i.e., if a polarization mode was too spectrally detuned, we would have not observed the two emission lines with polarizations associated with it). We note that the collection of horizontally polarized light is less efficient than that of the vertically polarized light due to a small spectral splitting of polarization modes of this particular cavity caused by fabrication imperfections. Despite such quantitative differences, the ability to access quantum dots with light beams orthogonal to each other is crucial for the realization of pulse sequences for coherently controlling the quantum dots spin for quantum information processing utilizing circularly polarized light \cite{Press2008,Stockill2016,Farfurnik2021}.

\begin{figure}[!h]
	\centering
			\includegraphics[width=0.7\textwidth]{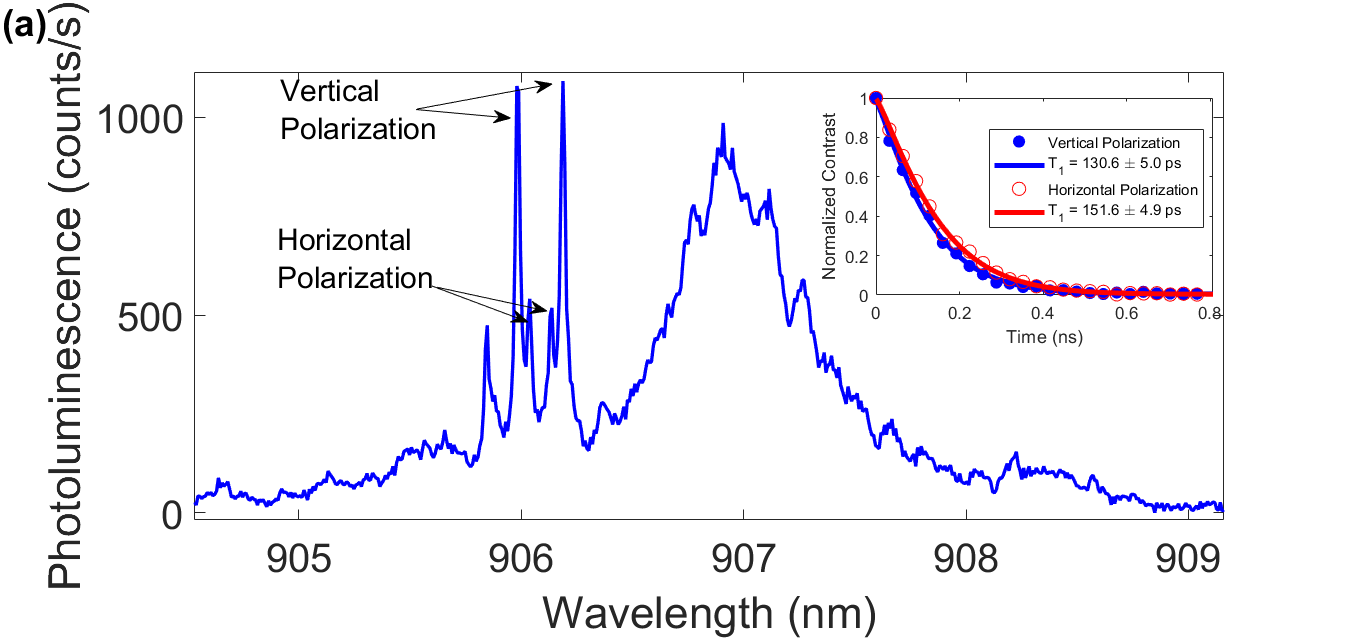}
					\includegraphics[width=0.7\textwidth]{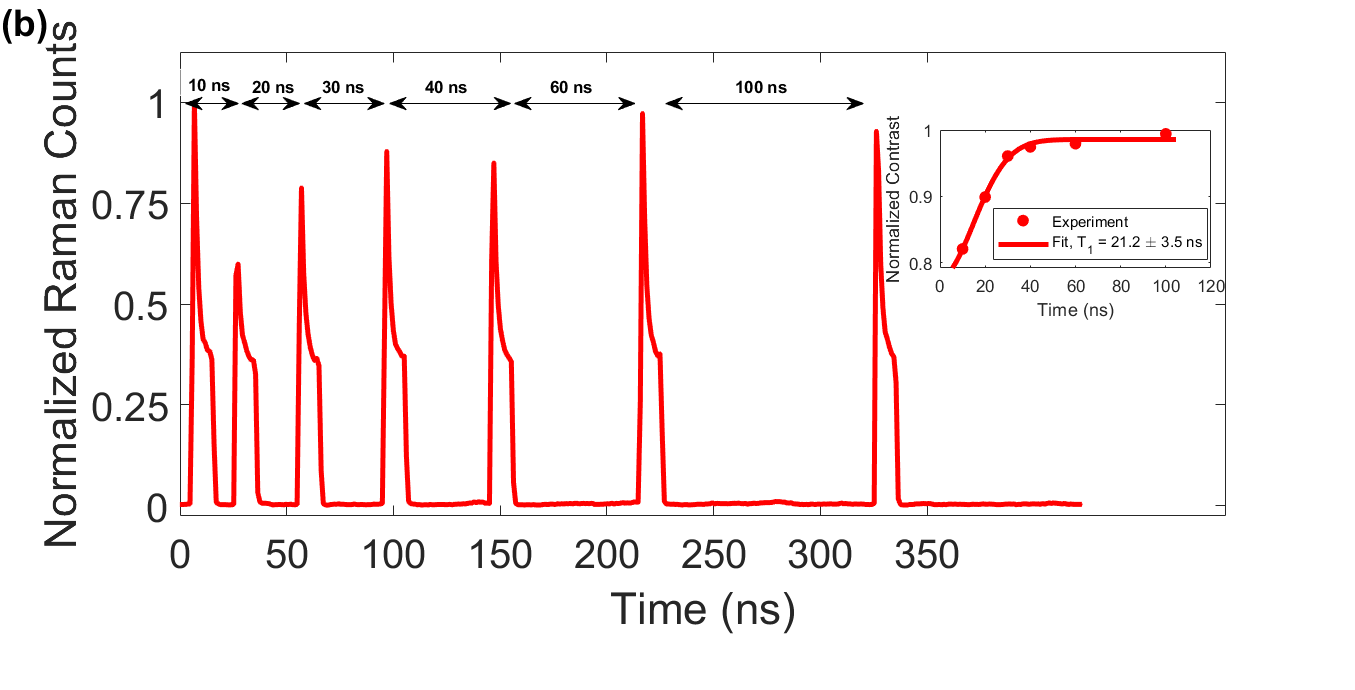}
	\caption{(a) The photoluminescence spectrum of an electrically charged quantum dot coupled to a bullseye cavity, under an external magnetic field of $B = 9$ T in the Voigt geometry and  ultrafast pulsed above-band excitations with an average optical power of 10 $\mu$W. The number of photons collected from the quantum dot is $\sim$ 25 times larger than the number of photons collected from dots in the bulk under $100$ $\mu$W above-band excitations. Inset: time-resolved measurements of the optical lifetimes of the dot via the differently polarized optical transitions. (b) Raman signals collected from the quantum dot under the same external magnetic field and resonant excitation pulses at varying times. The sharp peaks indicate the optical pumping of the spin, and the increasing heights of these peaks with time indicate the relaxation of the quantum dot spin. Inset: The total  photoluminescence signals collected under the application of the pulses (normalized by the photoluminescence induced by the first pulse), indicating spin relaxation on a timescale of $T_1 = 21.2 \pm 3.5$ ns.}
	\label{fig:fig4}
\end{figure}

To further emphasize the potential of controlling the quantum dot spin in the bullseye cavity, we use laser pulses resonant with one of the optical transitions of the dot to optically pump the spin (under $B = 9$ T). Varying the free evolution time between these pulses and measuring the emission of Raman signal from the dot results in the saturation behavior depicted in Fig. 4 (b). The sharp peak of the Raman signal emitted under the application of the first pulse indicates the optical initialization of the spin to one of its ground states. Then, the application of additional pumping pulses should not induce any Raman signal. Experimentally, however, the pulses induce undesired Raman signals that saturate for the free evolution time of $T_1=21.2 \pm 3.5$ ns [extracted from the least-square-fitting shown in the inset of Fig. 4(b)]. This saturated Raman signal represents the relaxation of the quantum dot spin, which reduces the spin initialization fidelity down to $\approx 25\%$. The observed spin relaxation is dominated by two physical mechanism. First, the main mechanism that causes spin relaxation is the co-tunneling of the electron confined in the dot with the electrons in the n-type back contact \cite{Gillard2021}, which results in our sample in spin relaxation times of few tens of ns even for dots in the bulk. These natural relaxation times can be further extended by orders of magnitude by modifying the tunnel barriers (GaAs layers) of the diode \cite{Lu2010,Gillard2021}. The second (minor) cause for the short spin relaxation time observed here is the spectral proximity of the bullseye cavity ($\sim$ 905 nm) to the wetting layer. Given this spectral proximity, pumping an optical transition of the dot coupled to the cavity may lead to a residual above-band pumping of both spin states, thereby reducing the spin initialization fidelities (by few additional percent) compared to the ones observed for dots with optical wavelengths of $\sim$ 930 nm. This residual pumping can be mitigated by designing and utilizing bullseye cavities with higher resonant wavelengths, e.g., by increasing the dimensions of the rings. The mitigation of both natural and laser-induced spin relaxation mechanisms could enable high fidelity spin control of the quantum dot spin using low laser powers, thereby upgrading the potential of these dots for quantum information processing. 

To conclude, nearly polarization-degenerate bullseye cavities in charge tunable devices can offer a spin-photon interface with an efficient optical access. By leveraging the low charge noise associated with a p-i-n-i-n diode, quantum dots exhibit single photon emission lifetimes as short as $80$ ps and lead to transparency in the cavity reflectivity of $\approx 80\%$. The cavities increase the efficiency of exciting and collecting photons from the dots by over an order of magnitude. While the optical lifetimes measured here could be further shortened by utilizing resonant pulse trains \cite{Liu2018}, the optical interface could be improved by utilizing a distributed Bragg reflector, and the fidelity of pumping the quantum dot spin can be improved by modifying the physical dimensions of the cavity and the diode. Combined with the nearly degenerate polarization modes provided by the bullseye  cavity, such improved optical interfaces could enable the coherent manipulation of quantum dot spin for quantum information processing and quantum sensing.

\subsection*{Supporting Information}
Sample growth procedures, fabrication details, description of the experimental apparatus, results of magnetic field scans indicating the presence of uncharged and charged quantum dots, and description of the theoretical model used for the simulation of dipole-induced transparency.
\begin{acknowledgement}
We thank Sascha Kolatschek and Robert Pettit for useful discussions. This work has been supported by the Physics Frontier Center at the Joint Quantum Institute, the National Science Foundation (Grants PHY1415485 and ECCS1508897), and the ARL Center for Distributed Quantum Information (Grant W911NF1520067). D. F.  acknowledges support by the Fulbright Postdoctoral Fellowship and the     Israel Council for Higher Education Quantum Science and Technology Scholarship. A.S.B. and S.G.C acknowledge support from the U.S Office of Naval Research.

\end{acknowledgement}
\bibliography{../../../../../mytex/mybibliography}
\end{document}